\documentclass{appolb}
\usepackage{graphicx}
\usepackage{amsmath}
\usepackage{url}
\begin{document}
\title{Single-step Quantum Simulation of Two Nucleons%
\thanks{Presented at Mazurian Lakes Conference on Physics (2025)}%
}
\author{Bhoomika Maheshwari
\address{Grand Accélérateur National d'Ions Lourds, CEA/DSM - CNRS/IN2P3, Bvd Henri Becquerel, F-14076, Caen, France}
\\[3mm]
{Paul Stevenson % of different affiliation
\address{School of Mathematics and Physics, University of Surrey, Guildford, Surrey GU2 7XH, United Kingdom}
}
\\[3mm]
P. Van Isacker
\address{Grand Accélérateur National d'Ions Lourds, CEA/DSM - CNRS/IN2P3, Bvd Henri Becquerel, F-14076, Caen, France}
}
\maketitle
\begin{abstract}
Quantum computing offers a scalable approach to solving the nuclear shell model, a highly complex and exponentially scaled many-body problem. This work presents a numerical simulation of the subspace search variational quantum eigensolver (SSVQE) combined with an adaptive derivative-assembles pseudo-trotter (ADAPT) ansatz to obtain the low-lying states of any nuclear system in a single optimization run. As an example, we apply this method in this work to a trivial identical nucleon system, two nucleons in the $0p_{3/2}$ orbital, mapped to 4 qubits depicting m-scheme single-particle states including a surface delta effective interaction using the Jordan-Wigner transformation. The ADAPT-SSVQE algorithm, by utilizing a symmetry-preserving double-excitation ADAPT operator pool, uniquely optimizes a weighted energy sum, forcing the simultaneous convergence of two lowest states within the total angular momentum $M_J=0$ subspace. We demonstrate the accuracy of the method by benchmarking against the exact diagonalization, confirming its potential for probing nuclear structure and pairing phenomena on current and near-future quantum devices without requiring multi-step procedure for excited states.   
\end{abstract}
  
\section{Introduction}
Pairing is a universal feature of quantum many-body systems manifesting in superconductivity, superfluidity, and nuclear structure. In nuclei, pairing correlations lower the ground-state energy and give rise to characteristic excitation spectra. The nuclear shell model provides a robust framework to study such correlations by representing nucleons as fermions moving in a mean-field potential with a residual interaction~\cite{smbook}. However, the exact diagonalization of shell-model Hamiltonians becomes intractable for large model spaces due to exponential scaling. Quantum algorithms such as the variational quantum eigensolver (VQE)~\cite{vqe} offer a path towards approximate solutions on near-term quantum devices~\cite{nisq}. While standard unitary coupled cluster (UCC) ansatzes~\cite{ucc} are general but costly; adaptive approaches (ADAPT-VQE)~\cite{adapt} construct ansatzes iteratively guided by physical gradients thus ensuring compactness. The quantum simulation of ground state solutions for the nuclear shell model has recently been approached using ADAPT-VQE~\cite{spain}. A more specific quantum simulation of shell model in $^{58}$Ni has been demonstrated using problem-specific ansatzes optimizing separately for ground and two excited states~\cite{bharti}.  

The basic variational method in quantum mechanics, upon which the VQE is based, targets only lowest-energy states within a Hilbert space.  In nuclear problems, one is often interested in a full or partial spectrum of excited states, and we are motivated in this direction, as others have been with VQE-inspired methods \cite{qsd,hobday,li}.

In this work, we present ADAPT-VQE with subspace-search (SSVQE~\cite{ssvqe}) to simulate shell model Hamiltonian for surface delta interaction for identical nucleons. We show the results for the example case of a single $0p_{3/2}$ orbital. This approach enables the simultaneous approximation of mutually orthogonal ground and excited states while respecting angular momentum symmetries in a single optimization run.       

\section{Theoretical Framework}

The effective nuclear Hamiltonian $H_{eff}$ is truncated to one- and two-body terms, represented in second quantization~\cite{smbook}:
\begin{equation}
    H_{eff}=\sum_{ab} \epsilon_{ab} a^\dagger_a a_b + \frac{1}{4} \sum_{abcd} \langle ab | V| cd \rangle_A a^\dagger_a a^\dagger_b a_d a_c 
\end{equation}
where $a^\dagger_i$ and $a_i$ are creation and annihilation operators for a m-scheme single-particle orbital $i$, $\epsilon_{ab}$ are one-body energies (which are set to zero in this simulation considering only relative spacing of the excited and ground states), and $\langle ab |V|cd\rangle_A $ are the anti-symmetrized two-body matrix elements. We employ a schematic surface delta interaction $V$ which yields $J$-coupled matrix elements $V_J$. The interaction strength $V_J$ is calculated from Clebsch-Gordan coefficients, $(j_a m_a j_b m_b | J M)$ for the two interacting particles in the same orbital as~\cite{smbook},
\begin{equation}
    V_J= \langle j^2; J | V| j^2 ; J\rangle =-\frac{(2j+1)^2V_1}{2(2J+1)} (j,-1/2,j,1/2|J,0)^2
\end{equation}
For identical nucleons (pp or nn), the Pauli principle enforces $V_J=0$ for odd $J$. We set $V_1=1$ MeV for this schematic calculation. The two-body matrix elements in m-scheme are then constructed by converting the coupled basis of $jj-$coupling into the uncoupled basis of $m-$scheme.  

The second-quantized Hamiltonian is mapped to a qubit operator using the Jordan-Wigner (JW) transformation~\cite{jw} using
\begin{eqnarray}
    a_j \rightarrow \Big[ \Pi_{k=1}^{j-1} \text{Z}_k \Big] \frac{1}{2} (\text{X}_j+ i \text{Y}_j) \\
    a^\dagger_j \rightarrow \Big[ \Pi_{k=1}^{j-1} \text{Z}_k \Big] \frac{1}{2} (\text{X}_j- i \text{Y}_j)   
\end{eqnarray}
where X, Y, Z are Pauli spin matrices acting on qubit $j$. The string of Z matrices in square brackets is essential for ensuring that operators on different qubits correctly anti-commute, maintaining the anti-symmetrization for a fermionic system. After this transformation, the Hamiltonian becomes a sum of Pauli strings which is a form that can be measured on a quantum computer. 

The total Hilbert space dimension for two protons in 4 $m$-scheme single-particle qubits corresponding to the $0p_{3/2}$ orbital is $\begin{pmatrix}
    4 \\ 2
\end{pmatrix}= 6$. In more realistic scenarios, we have multiple nucleons occupying higher $j$, or even many $j$ orbitals, leading to a larger $(2j+1)$, or $\sum(2j+1)$ $m$-scheme single-particle qubits and could reach the Hilbert space dimensions greater than $\sim{10}^{10}$, making classical exact diagonalization infeasible. 

\begin{figure}[!htb]
    \centering
    \includegraphics[width=0.75\textwidth]{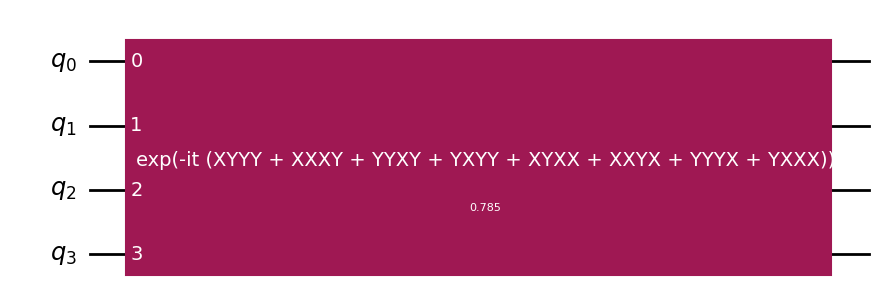}
    \caption{Quantum circuit of the ansatz, $U(\theta)=exp(-i\theta_0A_1)$, in terms of Pauli evolution gate in terms of double-excitation operator $A_1$ (shown here for brevity without coefficients of the involved terms) with $t\equiv \theta_0=0.785$.}
    \label{fig:qc}
\end{figure}

\begin{figure}[!htb]
    \centering
    \includegraphics[width=0.85\textwidth]{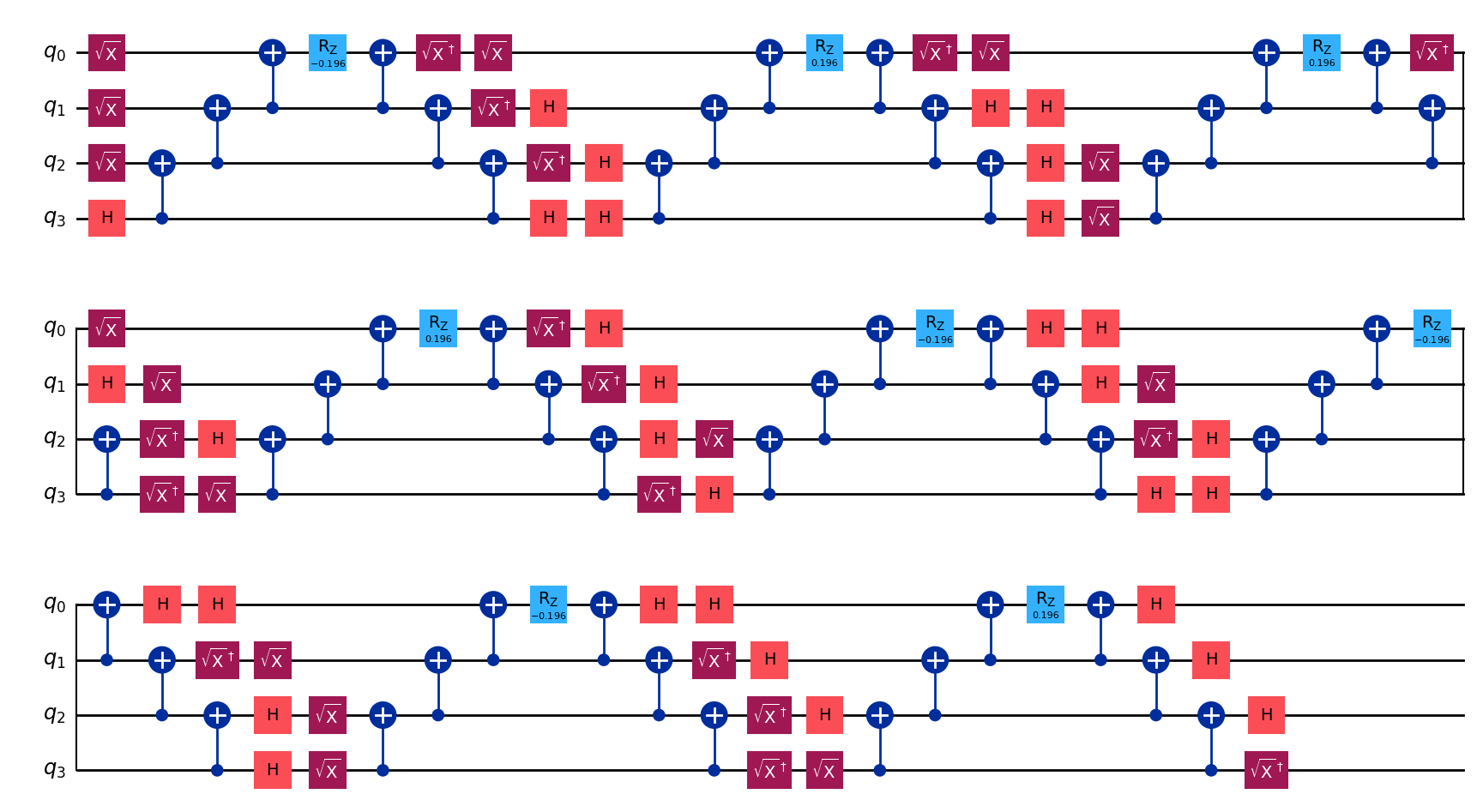}
    \caption{An equivalent representation of the quantum circuit shown in Fig.~\ref{fig:qc} but in terms of single and double-qubit gates suitable for the real hardware implementation.}
    \label{fig:qcd}
\end{figure}

\section{Variational Algorithm}

By leveraging the principles of superposition and entanglement, a quantum computer with $N$ qubits can naturally represent a state in a Hilbert space of dimension $2^N$. VQE~\cite{vqe} is based on the variational principle of quantum mechanics which states that the expectation value of the Hamiltonian for any trial wavefunction is always greater than or equal to the true ground state energy. It has emerged as a leading choice for utilizing noisy-intermediate-scale quantum (NISQ) devices~\cite{nisq}. VQE uses a quantum computer to prepare and measure a parameterized trial state $|\Psi(\vec\theta) \rangle = U(\vec\theta) | \phi_0 \rangle$ and a classical computer to optimize these parameters $\vec\theta$ that minimizes the energy expectation value $\langle \Psi(\vec\theta) | \hat{H} | \Psi(\vec\theta) \rangle$. $U(\vec\theta)$ is called the ansatz, that is, the quantum circuit by which the initial states are evolved, imposing the essential nucleonic correlations. The resulting quantum circuit is often expressed in terms of single and double-qubit gates. Examples of single-qubit gates are Pauli-x matrix X, Pauli-y matrix Y, Pauli-z matrix Z, Hadamard H, rotation with respect to x, y, z-axis in the Bloch sphere, $R_x,R_y,R_z$. %These rotation gates in the extreme limit of $\theta=\pi$ turn into X, Y, Z gates, respectively.
One of the most used double-qubit gates is CNOT (CX) gate, depicted as $\oplus$ and $\bullet$ connecting two qubits by a vertical line. The qubit with $\bullet$ is called control qubit while the qubit with $\oplus$ is called target qubit.  Further details of quantum gate can be found in textbooks \cite{wong}. 

Subspace Search Variational Quantum Eigensolver (SSVQE)~\cite{ssvqe} is an extension of the VQE to find multiple excited states simultaneously and was originally introduced in molecular simulations. We adopt it similarly for the nuclear physics in this work, proving its credibility for a simple $0p_{3/2}$ orbital (More details with realistic examples are to be followed in a forthcoming article). It replaces the minimization of a single state expectation value $\langle \Psi(\vec\theta) | \hat{H} | \Psi(\vec\theta) \rangle$ with a weighted sum of expectation values from multiple initial mutually orthogonal $k$ number of states, $|\Psi_i\rangle = U(\vec\theta) |\phi_i\rangle$ with a weighted loss defined by
\begin{eqnarray}
    L(\vec\theta) = \sum_{i=0}^{k-1} w_i \langle \phi_i | U^\dagger (\vec\theta) \hat{H} U(\vec\theta) | \phi_i \rangle  = \sum_{i=0}^{k-1} w_i E_{sorted,i}
\end{eqnarray}
where $w_i$ are descending positive weights (in this work, $w_i=k-i$) to prioritize the output states $|\Psi_i\rangle$ converging to the lowest $k$ energy eigenstates of $H$. The calculated energies are sorted at each optimization step to maintain the correct weighting order. 

Instead of a using a fixed circuit (like Unitary Coupled Cluster approaches~\cite{ucc}), ADAPT-VQE iteratively builds the ansatz $U(\vec\theta)$ by selecting the most relevant Pauli string operator from a predefined pool $P$. The calculation proceeds in a fixed $M_J=\sum_i m_i$ subspace, enforced by the ADAPT pool.  In this work, the pool is generated from all possible projection of total angular momentum $M_J$ conserving double-excitation operators, $T_{ab}^{cd}=a_c^\dagger a_d^\dagger a_b a_a$. This is converted to a Hermitian Pauli operator $A \in P$ suitable for exponential ansatz, $\exp(A)$, with $A=i(T-T^\dagger)$ where $T^\dagger$ is the complex conjugate of $T$. $T$ is chosen as a double-excitation fermionic operator involving two creation and two annihilation operators in this work. The use of a symmetry-constrained operator pool ensures the ansatz generates valid nuclear states and reduces the initial pool size.  On the other hand, it reduces the possible iterative paths to true ground states that are allowed by a more general symmetry-breaking pool \cite{tang}.  Limiting the pool in this way can only be justified \textit{a posteriori} if the results are successful for the present and future larger-scale studies.

At each iteration $iter$, the next operator $A_{iter}$ is chosen from $P$ by maximizing the weighted absolute gradient contribution $G_{iter}$ to the SSVQE cost function,
\begin{equation}
    G_{iter}={max}_{A\in P} \Bigg| \sum_{i=0}^{k-1} w_i \langle \Psi_i(\vec\theta) | i [H, A_{iter}] | \Psi_i(\vec\theta) \rangle \Bigg| 
\end{equation}
The ADAPT algorithm chooses the operator with maximum gradient to evolve the anstaz and proceeds until $G_{iter}$ falls below a defined tolerance, ${10}^{-2}$ in this simulation.  The final ADAPT ansatz is a product of Pauli evolution gates,
\begin{equation}
    U(\vec\theta)=\Pi_{iter=0}^{iter_{max}}exp(-i \vec\theta_{iter} A_{iter})
\end{equation}
which is unitary, so if the initial states are chosen appropriately with mutual orthogonality then the evolution using $U(\vec\theta)$ will strictly maintain their orthogonality and symmetries without any post-efforts. 

\section{Results and Discussion}

The calculation uses 4 qubits corresponding to two identical nucleons in a $0p_{3/2}$ orbital. The qubits $q_0, q_1, q_2, q_3$ correspond to $m_j=-3/2, -1/2$, 1/2, 3/2, respectively. Following Jordan-Wigner mapping, the qubit-Hamiltonian $H$ consists of the nineteen Pauli-strings of length 4, with five IIII, IIIZ, IIZI, IZII, ZIII one-body terms, and fourteen IIZZ, IZIZ, IZZI, ZIZI, ZZII, ZIIZ, (six terms with I and Z matrices), XYXY, XXYY, YYYY, YXXY, XYYX, XXXX, YYXX, YXYX (eight terms with X and Y matrices), two-body terms. The simulation targets the $M_J=0$ subspace. The initial orthogonal states $|\phi_i\rangle$ are constructed from the two lowest-energy Hartree-Fock basis states that satisfy symmetry constraints. The first state $|\phi_0 \rangle$ occupies qubits 0 and 3, $|1001\rangle$ and the second state $|\phi_1 \rangle$ occupies qubits 1 and 2, $|0110\rangle$. This choice provides an efficient initial reference for SSVQE optimization.

The ansatz is chosen adaptively, utilizing $M_J$ conserving double-excitation operators. In the present case, there would only be two such operators: The first one $A_0$ is trivial being IIII identity operator with a zero coefficient with which the gradient of Hamiltonian would always be zero, while the second operator $A_1$ is the only possible double excitation for this system in terms of X and Y matrices, combining  XYYY, XXXY, YYXY, YXYY, XYXX, XXYX, YYYX, and YXXX with respective coefficients following Jordan-Wigner mapping. It means that the algorithm must use the second double-excitation operator, $A_1$ to create an ansatz. The present calculation converges extremely rapidly in only two ADAPT iterations reaching a final gradient of $1.8\times{10^{-4}}$ below the tolerance. The gradient value of this second operator with the Hamiltonian turns out to be $1.6$ depicting that the initial Hartree-Fock states are very far from the true correlated eigen states. Those initial states are needed to be evolved using the ansatz $exp(-i \theta_0 A_1)$ with $\theta_0$ with the resulting quantum circuit shown in Fig.~\ref{fig:qc}. Note that in this simple example, a single parameter is enough to characterise the ansatz.  The classical optimization is performed using the COBYLA algorithm and $\theta_0$ is found to be 0.785. This means that $A_1$ operator perfectly captures all the correlations necessary to transition from the non-interacting Hartree-Fock configurations to the highly correlated $J=0$ and $J=2$ eigenstates. The calculated ground state energy $E_0=-2.0$ MeV and excited state energy $E_1=-0.4$ MeV have been benchmarked against the exact classical diagonalization results. 

We also present in Fig.~\ref{fig:qcd} the decomposed view of quantum circuit in terms of single-and double-qubit gates which could directly be supported by the real quantum hardware and is equivalent to the single Pauli evolution gate shown in Fig.~\ref{fig:qc}. The number of CNOT gates in this circuit is 48 and the circuit depth, defining the minimum number of sequential time steps to execute all the involved gates, is 72. The circuit width is $4$ while the number of single-qubit $\sqrt{\text{X}}$ and its conjugate $\sqrt{\text{X}^\dagger}$ gates is 32. The number of Hadamard gates H is also 32 which creates an entangled state, representing an equal superposition of two basis states. The eight number of rotation gates along $z-$ axis, $R_z$ are required owing to the eight terms in operator $A_1$.    

The simulation uses ``qiskit estimator" for noise-free calculation of expectation values. This primitive efficiently processes the multiple circuits and observables required for both the SSVQE cost function and ADAPT gradient calculation. %To conclude, ADAPT-SSVQE is a single-step spectroscopic variational quantum eigensolver for both ground and excited states, eliminating the need for complex multi-step post-processing or state-averaging techniques.  

\section{Conclusion}

We have successfully simulated the nuclear pairing problem of two identical nucleons in the $0p_{3/2}$ orbital on a noise-free quantum simulator. The nuclear shell model Hamiltonian with surface-delta interaction is mapped onto qubits with Jordan-Wigner mapping. The variational minimization ADAPT-SSVQE algorithm generates the two initial Hartee-Fock states and evolve them using a $M_J$ conserving double-excitation operator. For this foundational problem, the algorithm finds both the ground and excited states of the system within one iteration with a single $\theta$ parameter optimized classically. This work demonstrates that adaptive hybrid quantum-classical algorithms within a subspace search can efficiently capture correlations of nuclear many-body Hamiltonian. Future work will focus on realistic scenarios involving more qubits, and also incorporating both protons and neutrons along with isospin degree of freedom.  Preliminary results in this direction suggest that the method is able to calculate all the five allowed states for two-nucleons in $g_{9/2}$, $^{70}$Ni, within a single optimization run.    

\section*{Acknowledgments}
The author BM gratefully acknowledges the financial support from the HORIZON-MSCA-2023-PF-01 project, ISOON, under grant number 101150471. She also thanks Alahari Navin for useful and stimulating discussions. This work is also funded by UK STFC grant no. ST/Y000358/1.


\begin{thebibliography}{100}

\bibitem{smbook}
Shell Model Applications in Nuclear Spectroscopy,
P. J. Brussaard and P. M. W. Glaudemans
North-Holland Publishing Company (1977).

\bibitem{vqe}
McClean, J. R., Romero, J., Babbush, R. and Aspuru-Guzik, A. 
The theory of variational hybrid quantum-classical algorithms. 
New J. Phys. \textbf{18}, 023023 (2016). 
\url{https://doi.org/10.1088/1367-2630/18/2/023023}.

\bibitem{nisq}
Bharti, K. \textit{et al.} 
Noisy intermediate-scale quantum algorithms. 
Rev. Mod. Phys. \textbf{ 94}, 015004 (2022). 
\url{https://doi.org/10.1103/ RevModPhys.94.015004}.

\bibitem{ucc}
Anand, A. \textit{et al.} 
A quantum computing view on unitary coupled cluster theory. 
Chem. Soc. Rev. \textbf{51}, 1659–1684 (2022). 
\url{https://doi.org/10.1039/D1CS00932J}.

\bibitem{adapt}
Grimsley, H.R., Economou, S.E., Barnes, E. \textit{et al.} 
An adaptive variational algorithm for exact molecular simulations on a quantum computer. 
Nat Commun. \textbf{10}, 3007 (2019). 
\url{https://doi.org/10.1038/s41467-019-10988-2}

\bibitem{spain}
Pérez-Obiol, A., Romero, A.M., Menéndez, J. \textit{et al.} 
Nuclear shell-model simulation in digital quantum computers. 
Sci Rep \textbf{13}, 12291 (2023). 
\url{https://doi.org/10.1038/s41598-023-39263-7}

\bibitem{bharti}
Bhoy, B. and Stevenson, P.
Shell-model study of $^{58}$Ni using quantum computing algorithm.
New Journal of Physics, \textbf{26(7)}, 075001 (2024).
\url{https://doi.org/10.1088/1367-2630/ad5756}

\bibitem{qsd}
Zhang, J., and Lacroix, D.
Excited states from ADAPT-VQE convergence path in many-body problems: Application to nuclear pairing problem and $H_4$ molecule dissociation.
Phys. Lett. B \textbf{869}, 139841 (2025).
\url{https://doi.org/10.1016/j.physletb.2025.139841}

\bibitem{hobday}
Hobday, I. Stevenson, P. D., and Benstead, J. 
Variance minimization for nuclear structure on a quantum computer
Phys. Rev. C \textbf{111}, 064321 (2025).
\url{https://doi.org/10.1103/44k6-w3dt}

\bibitem{li}
Li, Ruo-Nan, Tau, Yuan-Hong, Liang, Jin-Min, Wu, Shu-Hui, and Fei, Shao-Ming
Full quantum eignesolvers based on variance
Phys. Scr. \textbf{99}, 095207 (2024)
\url{https://doi.org/10.1088/1402-4896/ad664c}

\bibitem{ssvqe}
Nakanishi, Ken M., Mitarai, K., and Fujii, K.
Subspace-search variational quantum eigensolver for excited states.
Phys. Rev. Research \textbf{1}, 033602 (2019).
\url{https://doi.org/10.1103/PhysRevResearch.1.033062}

\bibitem{jw}
Jordan, P. and Wigner, E. 
Über das Paulische Äquivalenzverbot, 
Zeitschrift für Physik \textbf{47 (9)}, 631 (1928).
\url{https://doi.org/10.1007/BF01331938}.

\bibitem{wong}
Wong, T.
Introduction to Classical and Quantum Computing, Rooted Grove, Omaha (2022)
\url{http://www.thomaswong.net}

\bibitem{tang}
Ho Lun Tang \textit{et al.}
Qubit-ADAPT-VQE: An Adaptive Algorithm for Constructing Hardware-Efficient Ansätze on a Quantum Processor
PRX Quantum 2, 020310 (2021)
\url{https://doi.org/10.1103/PRXQuantum.2.020310}

\end{thebibliography}
\end{document}